# Progress on Problem about Quantum Hamming Bound for Impure Quantum Codes

Zhuo Li and Lijuan Xing

*Abstract*—A famous open problem in the theory of quantum error-correcting codes is whether or not the parameters of an impure quantum code can violate the quantum Hamming bound for pure quantum codes. We partially solve this problem. We demonstrate that there exists a threshold $N(d,m)$ such that an arbitrary $((n,K,d))_m$ quantum code must obey the quantum Hamming bound whenever $n \geq N(d,m)$. We list some values of $N(d,m)$ for small $d$ and binary quantum codes.

*Index Terms*—Quantum Hamming bound, impure quantum codes, quantum error-correcting codes.

## I. Introduction

Quantum information can be protected by encoding it into a quantum error-correcting code. An $((n,K,d))_m$ quantum code is a $K$-dimensional subspace of the state space $\mathcal{H} = (\mathbb{C}^m)^{\otimes n}$ of $n$ quantum systems with $m$ levels that can detect all errors affecting less than $d$ quantum systems, but cannot detect some errors affecting $d$ quantum systems. A quantum code with a set of orthonormal basis $\{v_i\}$ is able to detect the error $E$ if and only if there exists a constant number $c_E$ such that

$$\langle v_i | E | v_j \rangle = c_E \delta_{ij}. \qquad (1)$$

An $((n,K,d))_m$ quantum code is said to be pure if and only if the constant $c_E$ in (1) is equal to 0 for all errors affecting less than $d$ quantum systems; otherwise, it is called impure.

In the theory of quantum error-correcting codes there is a well known bound on the pure quantum codes, i.e., the quantum Hamming bound which says an $((n,K,d))_m$ pure quantum code satisfies

$$K \sum_{i=0}^{\lfloor \frac{d-1}{2} \rfloor} (m^2-1)^i \binom{n}{i} \leq m^n. \qquad (2)$$

The quantum Hamming bound applies only to the pure quantum codes, which naturally poses a problem about whether an impure $((n,K,d))_m$ quantum code violating the Hamming bound (2) might exist [1]. The standard proof of (2) by a simple counting argument can fail for the impure quantum codes, fueling the interest in this problem. Since the occurrence of this problem, researchers have been trying to solve it. To this date this problem remains to be fully settled.

Gottesman showed in [2] that single and double error-correcting binary stabilizer codes cannot beat the quantum Hamming bound. Ashikhmin and Litsyn [3] showed that asymptotically binary quantum codes obey the quantum Hamming bound. More recently, Gottesman's result was generalized for nonbinary codes with distance three [4]. It can be seen that all these results are only applicable to stabilizer codes or binary codes. Such results are not known for general quantum codes yet.

In this paper, we make some progress on this problem. We pay our attention to the most general quantum codes and show that the results mentioned above are valid for general quantum codes. More importantly, we demonstrate that given $d$ and $m$, there exists a positive integer $N(d,m)$ such that when $n \geq N(d,m)$ the quantum code $((n,K,d))_m$ cannot beat the quantum Hamming bound.

## II. Key Result

Our approach is based on the idea introduced by Delsarte for classical codes in [5]. This idea has been used for binary quantum codes [3] and stabilizer codes [4] respectively. Here we extend it to the most general quantum codes.

**Theorem 1**. Let $\mathcal{C}$ be an $((n,K,d))_m$ quantum code. Suppose that $S$ is a nonempty subset of $\{0,\ldots,d-1\}$ and $T = \{0,\ldots,n\}$. Let $f(x) = \sum_{i=0}^{n} f_i P_i(x)$ be a polynomial satisfying the conditions
1) $f_t > 0$ for all $t$ in $S$, and $f_t \geq 0$ otherwise;
2) $f(t) \leq 0$ for all $t$ in $T \setminus S$,

where $P_i(x)$ is a Krawtchouk polynomial. Then

$$K \leq \frac{1}{m^n} \max_{t \in S} \frac{f(t)}{f_t}.$$

Zhuo Li and Lijuan Xing are with the State Key Laboratory of Integrated Service Networks, Xidian University, Xi'an, Shannxi 710071, China (e-mail: lizhuo@xidian.edu.cn).

*Proof.* Using the universal framework of quantum codes, let $\{A_i\}$ and $\{A_i'\}$ be the Hamming weight distributions and the dual Hamming weight distributions of the quantum code $\mathcal{C}$ respectively. Then they are related by

$$\sum_{i=0}^{n} A_i' x^{n-i} y^i = \frac{K}{m^n} \sum_{r=0}^{n} A_r (x+(m^2-1)y)^{n-r} (x-y)^r \quad (3)$$

and $A_i = A_i'$ for all $i = 0, \ldots, d-1$, see [6]. If we write

$$x^{n-i} y^i = \left( \frac{(x+(m^2-1)y) + (m^2-1)(x-y)}{m^2} \right)^{n-i}$$

$$\cdot \left( \frac{(x+(m^2-1)y) - (x-y)}{m^2} \right)^i$$

$$= \frac{1}{m^{2n}} \sum_{r=0}^{n} P_r(i)(x+(m^2-1)y)^{n-r}(x-y)^r$$

then from (3)

$$A_r = \frac{1}{Km^n} \sum_{i=0}^{n} A_i' P_r(i).$$

Using these relations, we find that

$$Km^n \sum_{t \in S} f_t A_t \le Km^n \sum_{r=0}^{n} f_r A_r = Km^n \sum_{r=0}^{n} f_r \frac{1}{Km^n} \sum_{i=0}^{n} A_i' P_r(i)$$

$$= \sum_{i=0}^{n} A_i' \sum_{r=0}^{n} f_r P_r(i) = \sum_{i=0}^{n} A_i' f(i) \le \sum_{t \in S} A_t' f(t) = \sum_{t \in S} A_t f(t).$$

Thus

$$K \le \frac{1}{m^n} \left( \sum_{t \in S} A_t f(t) \right) / \left( \sum_{t \in S} f_t A_t \right) \le \frac{1}{m^n} \max_{t \in S} \frac{f(t)}{f_t}.$$

This theorem is convenient when one wants to find bounds by hand. In particular, any function $f$ satisfying the constraints of the theorem will yield a bound on quantum codes. In what follows we present the key result of this paper derived from Theorem 1.

Let $e = \lfloor \frac{d-1}{2} \rfloor$ and $\gamma = m^2 - 1$. Take $f_t = \left( \sum_{i=0}^{e} P_i(t) \right)^2$ and $f(x) = \sum_{r=0}^{n} f_r P_r(x)$. Notice that

$$P_i(x) P_j(x) = \sum_{k=0}^{n} P_k(x) \sum_{s=0}^{n-k} \binom{k}{2k+2s-i-j}$$

$$\cdot \binom{n-k}{s} \binom{2k+2s-i-j}{k+s-j} (\gamma-1)^{i+j-2s-k} \gamma^s,$$

a straightforward generalization of a similar expression in the binary case (see e.g. [7]). Hence

$$f_t = \sum_{i=0}^{e} \sum_{j=0}^{e} \sum_{k=0}^{n} P_k(t) \sum_{s=0}^{n-k} \binom{k}{2k+2s-i-j}$$

$$\cdot \binom{n-k}{s} \binom{2k+2s-i-j}{k+s-j} (\gamma-1)^{i+j-2s-k} \gamma^s.$$

This yields

$$f(t) = \sum_{r=0}^{n} \sum_{i=0}^{e} \sum_{j=0}^{e} \sum_{k=0}^{n} P_k(r) \sum_{s=0}^{n-k} \binom{k}{2k+2s-i-j}$$

$$\cdot \binom{n-k}{s} \binom{2k+2s-i-j}{k+s-j} (\gamma-1)^{i+j-2s-k} \gamma^s P_r(t)$$

$$= \sum_{i=0}^{e} \sum_{j=0}^{e} \sum_{k=0}^{n} \sum_{s=0}^{n-k} \binom{k}{2k+2s-i-j} \binom{n-k}{s}$$

$$\cdot \binom{2k+2s-i-j}{k+s-j} (\gamma-1)^{i+j-2s-k} \gamma^s \sum_{r=0}^{n} P_k(r) P_r(t)$$

$$= m^{2n} \sum_{i=0}^{e} \sum_{j=0}^{e} \sum_{s=0}^{n-t} \binom{t}{2t+2s-i-j} \binom{n-t}{s}$$

$$\cdot \binom{2t+2s-i-j}{t+s-j} (\gamma-1)^{i+j-2s-t} \gamma^s \quad (4)$$

where in the last step we use the relation $\sum_{r=0}^{n} P_k(r) P_r(t) = m^{2n} \delta_{k,t}$, see [8]. On the other hand, using the relation $P_0(x;n) + P_1(x;n) + \cdots + P_e(x;n) = P_e(x-1;n-1)$, see [8], we get

$$f_t = \left( \sum_{i=0}^{e} P_i(t) \right)^2 = \left( P_e(t-1;n-1) \right)^2$$

$$= \left( \sum_{j=0}^{e} (-1)^j \gamma^{e-j} \binom{t-1}{j} \binom{n-t}{e-j} \right)^2. \quad (5)$$

Then for fixed $d$ and $m$, we can express $f_t$ in the form $f_t = (\gamma^e/e!)^2 n^{2e} + o(n^{2e})$ for $t = 0, \ldots, d-1$ from (5), and $f(t)$ as $f(0) = m^{2n} \left( (\gamma^e/e!) n^e + o(n^e) \right)$, $f(t) = m^{2n} o(n^e)$ for $t = 1, \ldots, d-1$ and $f(t) = 0$ for $t = d, \ldots, n$ from (4), where $o(f(n))$ means $o(f(n))/f(n) \to 0$ as $n \to \infty$. Thus, in Theorem 1 if set $S = \{0, \ldots, d-1\}$, then $f(x)$ satisfies the conditions 1) and 2), and there exists a positive integer, say $N(d,m)$, such that when $n \ge N(d,m)$,

$$\max_{t \in S} \frac{f(t)}{f_t} = \frac{f(0)}{f_0}$$

$$= m^{2n} \sum_{s=0}^{e} \gamma^s \binom{n}{s} / \left( \sum_{j=0}^{e} \gamma^j \binom{n}{j} \right)^2 = m^{2n} / \sum_{j=0}^{e} \gamma^j \binom{n}{j}.$$

We have thus proved the following theorem.

**Theorem 2**. Given $d$ and $m$, there exists a positive integer $N(d,m)$ such that for all $n \geq N(d,m)$, if a quantum code $((n,K,d))_m$ exists it obeys the quantum Hamming bound.

## III. DISCUSSION

In fact, from (4) we can see that $f(t) = 0$ provided $t > 2e$. Thus when $d = 2e+2$, if instead we choose $S = \{0,\ldots,d-2\}$ in the above proof, Theorem 1 applies as well. This is to say for $d = 2e+1$ and $d = 2e+2$, we can choose identical $f(x)$ and $S$ in Theorem 1. Therefore we have the following property about $N(d,m)$.

**Theorem 3**. $N(2e+2,m) = N(2e+1,m)$.

This theorem tells us it is enough to find out the $N(d,m)$ with odd $d$. For example, TABLE I gives the values of $N(d,m)$ for binary quantum codes and small $d$. The case of $d = 1$ is trivial. $N(3,2) = 5$ implies that an $((n,K,3))_2$ quantum code obeys the quantum Hamming bound for all $n \geq 5$. But from the quantum Singleton bound, it follows that this claim is valid for $n < 5$ as well. Thus arbitrary $((n,K,3))_2$ quantum code obeys the quantum Hamming bound. Similarly, we can see from TABLE I that an $((n,K,5))_2$ cannot beat the quantum Hamming bound. In fact, using Theorem 1 it is not difficult to show that both claims are also valid for nonbinary quantum codes. Thus we can conclude that arbitrary single and double error-correcting codes cannot beat the quantum Hamming bound.

TABLE I
VALUES OF $N(d,m)$ FOR $m = 2$ AND SMALL $d$

| $d$ | 1 | 3 | 5 | 7 | 9 | 11 | 13 | 15 |
|---|---|---|---|---|---|---|---|---|
| $N(d,2)$ | 1 | 5 | 9 | 14 | 20 | 25 | 30 | 35 |

So far we have shown all our results. In this paper we focus on the famous open problem in the theory of quantum error-correcting codes: whether a quantum code violating the quantum Hamming bound might exist. We have solved this problem for long codes. Specifically, we have demonstrated that for arbitrary general quantum code there exists a threshold dependent only on the minimum distance and the level of the code such that the code cannot beat the quantum Hamming bound unless the code length is less than the threshold. This is a progress on this open problem. We have effectively reduced the room in which a quantum code might beat the quantum Hamming bound.